\documentclass[lettersize,journal]{IEEEtran}
\usepackage{bm}
\usepackage{subcaption}
\usepackage{booktabs}
\usepackage{float}
\usepackage{hvfloat}
\usepackage{amsmath,amsfonts}
\usepackage{algpseudocode}
\usepackage{algorithm}
\usepackage{array}
\usepackage{multirow}
\usepackage{textcomp}
\usepackage{stfloats}
\usepackage{url}
\usepackage{verbatim}
\usepackage{graphicx}
\usepackage{cite}
\usepackage{graphicx}
\usepackage{lipsum}
\usepackage{caption}
\usepackage{rotating}
\usepackage{caption}
\usepackage{lipsum}
\usepackage{eqparbox}
\usepackage{etoolbox}  
\usepackage{xcolor}
\usepackage{cuted}

\begin{document}

\begin{titlepage}
    \vspace*{\fill}
        \large
        © 2024 IEEE.  Personal use of this material is permitted.  Permission from IEEE must be obtained for all other uses, in any current or future media, including reprinting/republishing this material for advertising or promotional purposes, creating new collective works, for resale or redistribution to servers or lists, or reuse of any copyrighted component of this work in other works.
    \vspace*{\fill}
\end{titlepage}

\newpage
\title{Uncertainty Distribution Assessment of Jiles-Atherton Parameter Estimation \\ for Inrush Current Studies}

\author{Jone Ugarte Valdivielso, Jose I. Aizpurua Unanue,~\IEEEmembership{Senior Member,~IEEE} and Manex Barrenetxea Iñarra 
	\thanks{J. Ugarte, J. Aizpurua \& M. Barrenetxea are with the Electronics \& Computing Department, Mondragon Unibertsitatea, Arrasate, Spain; J. Aizpurua is also with Ikerbasque, Basque Foundation for Science, Bilbao, Spain.}
}

\markboth{IEEE Transaction on Power Delivery}%
{Shell \MakeLowercase{\textit{et al.}}: Bare Demo of IEEEtran.cls for IEEE Journals}

\maketitle

\begin{abstract} 
Transformers are one of the key assets in AC distribution grids and renewable power integration. During transformer energization inrush currents appear, which lead to transformer degradation and can cause grid instability events. These inrush currents are a consequence of the transformer's magnetic core saturation during its connection to the grid. Transformer cores are normally modelled by the Jiles-Atherton (JA) model which contains five parameters. These parameters can be estimated by metaheuristic-based search algorithms. The parameter initialization of these algorithms plays an important role in the algorithm convergence. The most popular strategy used for JA parameter initialization is a random uniform distribution. However, techniques such as parameter initialization by Probability Density Functions (PDFs) have shown to improve accuracy over random methods. In this context, this research work presents a framework to assess the impact of different parameter initialization strategies on the performance of the JA parameter estimation for inrush current studies. Depending on available data and expert knowledge, uncertainty levels are modelled with different PDFs. Moreover, three different metaheuristic-search algorithms are employed on two different core materials and their accuracy and computational time are compared. Results show an improvement in the accuracy and computational time of the metaheuristic-based algorithms when PDF parameter initialization is used.

\end{abstract}


\vspace{-0.5cm}

\section{Introduction}
\label{sec:Intro}

\begin{figure}[!h]
     \vspace{-5pt}
	\centering
	\includegraphics[width=0.8\columnwidth]{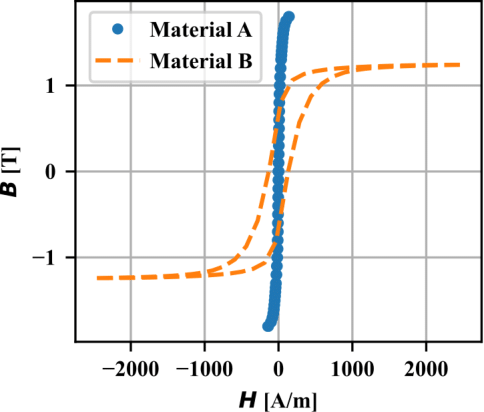}
	\caption{B-H curves for the two materials analysed in this study.}
	\label{fig:BH_ini}
    \vspace{-10pt}
 \end{figure}

\IEEEPARstart{T}{ransformers} are key components for the reliable and efficient operation of the grid. During energization, they suffer from inrush currents, which affect their reliability and life expectancy. Inrush current is a transient phenomenon that can affect the grid's stability \cite{Brunke2001}. In order to minimize these effects, the inrush current has to be reduced. To do so, several solutions can be adopted. Existing inrush minimization techniques include the oversizing of the transformer, addition of a pre-insertion resistor, pre-magnetization of the transformer or controlled switching performed by a Circuit Breaker (CB) \cite{Alassi2022}, \cite{Brunke2001_2}. Controlled switching is the most commonly used strategy, where the opening and closing times of the CB are controlled \cite{Cano2015}. The connection time depends on the remanent flux of the transformer's core at the transformer connection instant. Furthermore, the inrush current magnitude depends on the magnetic saturation level of the core during energization, which is described by its B-H curve (see example in Fig.~\ref{fig:BH_ini} for the B-H curves of the materials used in this study). In this context, an accurate model of the core is advantageous to avoid improper switching actions and limit the impact of the inrush current on the transformer health \cite{Chiesa2010}.

The relation between the B-H curve and the inrush current can be comprehended through basic electromagnetic principles. The flux density [$B$] can be obtained from the voltage on the primary side of the transformer and Faraday’s law of electromagnetic induction. The electromagnetic field strength [$H$] is derived from the hysteresis curve of the core material. Then, the magnetizing current of the transformer can be directly calculated through Ampère’s circuit law \cite{Jiles1991_4}.

\subsection{Related Work}

The non-linear dynamic behaviour of the transformer core is normally represented by the Jiles-Atherton (JA) model due to its better efficiency compared to other models such as the Preisach model \cite{Chiesa2010}. The JA model solves a Partial Differential Equation (PDE) through an equivalent method, which requires five parameters related to the physical properties of the core material. However, the accurate estimation of these parameters is challenging \cite{Liorzou2000}. The simplest technique to estimate JA parameters is a trial-and-error brute-force algorithm \cite{Sarker2022}. In this algorithm, the entire parameter range is swept, and consequently, the computational time is very high. Therefore, alternative search techniques such as metaheuristic-based algorithms are preferred over trial-and-error due to their shorter computational time and efficient search of the global minimum.

In this setting, the most used metaheuristic evolutionary techniques in literature are Genetic Algorithms (GA) \cite{Chwastek2006}, Particle Swarm Optimization (PSO) \cite{Coelho2012} and Differential Evolution (DE) \cite{Toman2008}. An extensive review of GA, PSO and DE algorithms in JA parameter estimation studies is presented in \cite{Quaranta2020}. 

Most of the presented studies disregard the initialization stage of the metaheuristic-search algorithm and use a random uniform JA parameter initialization strategy with default settings \cite{Rubežić2018, Naghizadeh2012, Zaman2015, Marion2008}. Nonetheless, the selection of inappropriate initial values and limits for JA parameters can lead to convergence to a local minimum or non-convergence, and high computational time. In this context, the convergence of PSO is shown to improve if the initialization is done in two stages, first by finding the best parameters in a set number of iterations and then by initializing the algorithm with those parameters \cite{Knypiński2012}. Additionally, a clustering approach based on self-organizing maps has been used to estimate the initial JA parameter values for a GA algorithm \cite{Salvini2002}. Besides, expressions for JA parameters have been derived to approximate their real values \cite{Jiles1991_2}. 

Alternatively, an extensive research study compares the use of different Probability Density Functions (PDFs) for the initialization of various metaheuristic algorithms in basic problems. Results show that using different PDFs, rather than uniform distribution, enhances the accuracy of the algorithm. Moreover, the findings indicate that PSO's accuracy is more sensitive to the initialization strategy than DE and GA. This implies that the choice of initialization strategy has a greater impact on the results for PSO than for the rest of the algorithms. However, the study states that depending on the complexity of the problem this sensitivity may change. For instance, the GA parameter initialization is shown to be more sensitive to complex problems \cite{Li2020}. According to another study where PSO parameter initialization with different PDFs is presented, the convergence speed is normally higher for Gaussian PDF than for uniform PDF \cite{Cazzaniga2015}. Yet, the accuracy of the results for these two initialization strategies is reported to be similar in most of the cases \cite{Cazzaniga2015}.

\subsection{Gap Identification, Contribution \& Impact}

Even though PDF parameter initialization can improve the accuracy of metaheuristic-based algorithms \cite{Li2020}, it is yet to be analysed how the influence of different initialization strategies, including available information or knowledge as equivalent PDFs, and propagation through the selected metaheuristic-search techniques, can affect the performance of the search-based optimization algorithms for JA parameter estimation. 

For this analysis, expert knowledge is required to assess the mean and the variance or uncertainty of each PDF. An expert, with extensive field knowledge, would confidently select initial parameters. Accordingly, a low-uncertainty PDF would reflect the initial value of parameters and the search area would be reduced. However, if the expert is wrong, the real value may be different from the expected value, and the algorithm will not converge to the minimum error. In contrast, a novel engineer without experience, may not be confident in the initial value parameters and may be interested in testing and searching for the effect of different initial values on accuracy and computational time to iterate and reduce the search space and design fast and efficient optimization algorithms. 

Therefore, it is necessary to devise an uncertainty-aware JA parameter estimation approach, which can investigate the impact of uncertainties associated with parameter initialization to improve the compromise between the error and the computational time. In this context, the main contribution of this research work is an uncertainty-aware JA parameter estimation approach based on metaheuristic-search algorithms. The framework has been tested with three different algorithms and validated with two different transformer core materials.

Different metaheuristic-search based algorithms are compared to a base-case, where brute-force has been employed to estimate the JA parameters. The parameter initialization for the metaheuristic algorithms has been done with different levels of uncertainty, which reflect available expert knowledge and previous data, by employing different PDFs. This information has been propagated through search-based algorithms and results show that using a Normal PDF for JA parameter initialization, instead of uniform parameter initialization, improves the JA parameter estimation accuracy and computational-efficiency.

The proposed parameter selection framework supports and improves the JA-based hysteresis modelling approach through an uncertainty-aware parameter initialization strategy. Thus, a better electromagnetic model of the transformer core is obtained.

In this context, core modelling is the first step for developing transformer models for transient studies. Generally, the duality-based transformer model is used for inrush current studies \cite{Chiesa2010, Chiesa2010_2}, which, in addition to other parameters, considers the hysteresis of both the transformer legs and yokes. This makes it mandatory to have accurate core models, otherwise errors would propagate up to the transformer transient model. Therefore, this research work has a direct impact on transformer transient studies because it enables the design of an accurate transformer core model for inrush minimization analysis. 

\subsection{Organization}

The rest of the paper is organized as follows. Section \ref{sec:proble_and_proposed_approach} introduces the basics of the theory behind the JA model, the fundamental of the chosen metaheuristic algorithms, the problem statement and the proposed approach for estimating JA parameters. Then, Section \ref{sec:case_study} presents the analysed materials, the base-case and the study cases. Section \ref{sec:numerical_results} covers the results obtained for the different cases and materials. In Section \ref{sec:discussion} the study is discussed followed by the conclusions in Section \ref{sec:Conclusions}.

\section{Problem Statement and Proposed Approach}
\label{sec:proble_and_proposed_approach}

\subsection{Basics of Jiles-Atherton}
\label{ssc:Jiles_Atherton_Review}
The Jiles-Atherton hysteresis model assumes that the total magnetization of a ferromagnetic material is the contribution of reversible and irreversible magnetization parts \cite{Jiles1984, Jiles1986}

\begin{equation}\label{eq:M_tot}
M = M_{\text{rev}}+M_{\text{irr}}
\end{equation} 

\noindent where $M$ is the total magnetization of the ferromagnetic material in [A/m] and $M_{\text{rev}}$ and $M_{\text{irr}}$ are respectively the reversible and irreversible parts, both in [A/m]. The irreversible term is obtained by using:
 
\begin{equation}\label{eq:M_irr}
M_{\text{irr}} = M_{\text{an}} - k\delta\frac{\text{d}M_{\text{irr}}}{\text{d}H_{\text{e}}}
\end{equation} 

\noindent where $M_{\text{an}}$ is the anhysteretic magnetization curve in [A/m], $k$ is the pinning parameter in [A/m], $\delta$ is a direction parameter which assumes a value of +1 if $\text{d}H/\text{d}t > 0$ and -1 if $\text{d}H/\text{d}t < 0$, and $H_{\text{e}}$ is the effective field in [A/m] calculated by:

\begin{equation}\label{eq:H_e}
H_{\text{e}} = H+ \alpha M
\end{equation} 

\noindent where $H$ is the magnetic field strength in [A/m] and $\alpha$ is unitless and accounts for the inter-domain coupling. The reversible magnetization is calculated as follows:

\begin{equation}\label{eq:M_rev}
M_{\text{rev}} = c(M_{\text{an}} - M_{\text{irr}})
\end{equation} 

\noindent where $c$ is unitless and represents the coefficient of proportionality and $M_{\text{an}}$ is the anhysteretic magnetization given in (\ref{eq:Langevin}). Anhysteretic magnetization refers to the ideal magnetization curve without hysteresis. A Langevin function is used to obtain the anhysteretic magnetization, defined as follows \cite{Jiles1991}:

\begin{equation}\label{eq:Langevin}
M_{\text{an}} = M_{\text{s}}\left [\coth{\left(\frac{H_{\text{e}}}{a}\right)}-\frac{a}{H_{\text{e}}} \right]
\end{equation} 

\noindent where $M_{\text{s}}$ is the saturation magnetization in [A/m] and $a$ quantifies the domain wall density in [A/m]. However, other expressions as the models given by Frölich \cite{Chiesa2010, Mork2007} and Annakkage \cite{Annakkage2000, Chandrasena2004} have also been used in literature to calculate this curve.

The Jiles-Atherton model is solved by a PDE that represents the total magnetization susceptibility, which is expressed as follows \cite{Sadowski2003}:

\begin{equation}\label{eq:Inverse_JA}
\frac{\text{d}M}{\text{d}H} = \frac{(1-c)\frac{\text{d}M_{\text{irr}}}{\text{d}H_{\text{e}}}+c\frac{\text{d}M_{\text{an}}}{\text{d}H_{\text{e}}}}{1-\alpha c \frac{\text{d}M_{\text{an}}}{\text{d}H_{\text{e}}}-\alpha (1-c)\frac{\text{d}M_{\text{irr}}}{\text{d}H_{\text{e}}}}
\end{equation} 

\noindent where $\frac{\text{d}M_{\text{irr}}}{\text{d}H_{\text{e}}}$ is obtained by deriving (\ref{eq:M_irr}) with respect to $H_{\text{e}}$ and is expressed as follows:

\begin{equation}\label{eq:M_irr_derived}
\frac{\text{d}M_{\text{irr}}}{\text{d}H_{\text{e}}} = \frac{(M_{\text{an}}-M_{\text{irr}})}{k\delta}
\end{equation}

Additionally, $\frac{\text{d}M_{\text{an}}}{\text{d}H_{\text{e}}}$ is obtained by deriving (\ref{eq:Langevin}) with respect to $H_{\text{e}}$ and is given as:

\begin{equation}\label{eq:Langevin_derived}
\frac{\text{d}M_{\text{an}}}{\text{d}H_{\text{e}}} = \frac{M_{\text{s}}}{a} \left[ 1 - \coth^2{\left(\frac{H_{\text{e}}}{a}\right)}-\left(\frac{a}{H_{\text{e}}}\right)^2 \right]
\end{equation} 

The B-H curve is solved by calculating the value of the magnetization in each time step $\Delta t$ of $H$:

\begin{equation}\label{eq:M_cal}
M(t+\Delta t) = M(t) + \frac{\text{d}M}{\text{d}H} \Delta H
\end{equation} 

\noindent where $t$ is the time step, $\Delta t$ is a discrete increment of $t$, $M(t+\Delta t)$ is the magnetization value at the instant $t+\Delta t$ in [A/m], $M(t)$ is the magnetization at $t$ in [A/m] and $\Delta H$ is the difference in the discrete increment of the field strength in [A/m].

Generally, the results are preferred to be presented in terms of flux density. The equation that relates the magnetization and flux density of the material, known as the \textit{Sommerfeld} convention, is given as $B = \mu_0 (H+M)$ \cite{Jiles1991_3}, where $\mu_0$ is the permeability of vacuum with a value of 4$\pi$x10$^{- 7}$ H/m. Therefore, the flux density in each time step is obtained as:

\begin{equation}\label{eq:B_cal}
B(t+ \Delta t) = \mu_0 \left[H(t+\Delta t) + M(t+\Delta t)\right]
\end{equation} 

\noindent where $B(t+ \Delta t)$ is the magnetic flux density in [T] at the instant $t+\Delta t$ and $H(t+\Delta t)$ is the field strength at the next step in [A/m].

\subsection{Basics of Metaheuristic Search Algorithms}
\label{ssc:Metaheuristics}

The selected algorithms are GA, PSO and DE due to their good accuracy and low computational complexity for the estimation of JA parameters \cite{Quaranta2020}.

GA uses the concept of human reproduction and mutation to optimize the solution search space. This algorithm initializes the population by a set of possible solutions. Then, candidate solutions (\textit{parents}) are selected and combined to produce a chromosome (\textit{child}) through crossover and mutation operations. By doing so, the algorithm creates better solutions. The main parameters needed for GA are population size, $\beta$ (integer for calculating the probability for each \textit{parent}'s selection), mutation step size and mutation rate \cite{Leite2004}.

PSO is a stochastic optimization process based on the social behaviour of a swarm. The population, named as particles, are scattered in a search space. Each particle has a position and a velocity, which changes in each iteration. The particles remember their best individual position from past iterations and also the position of the leader (best solution). Moreover, the swarm also shares the information with the rest of the particles accelerating the process. The main parameters needed for PSO are population size and initial particle velocity \cite{Knypiński2012}.    

DE is an evolutionary algorithm that optimizes the solution, changing and combining the existing population of individuals. For each individual a mutant is generated combining three randomly chosen individuals. The targeted individual and the mutant are combined through a crossover operation, which produces a trial. If this trial is better than the unchanged individual, the trial replaces its position. The main parameters for DE are population size and mutation rate and factor \cite{Toman2008}.

The parameters for each algorithm have been tuned to obtain optimal solutions with respect to accuracy and computational time, named as \textit{params} in Algorithm \ref{al:pseudocode1}.

\begin{figure*}[!t]
	\centering
    \vspace{-0.2cm}
	\includegraphics[width=2\columnwidth]{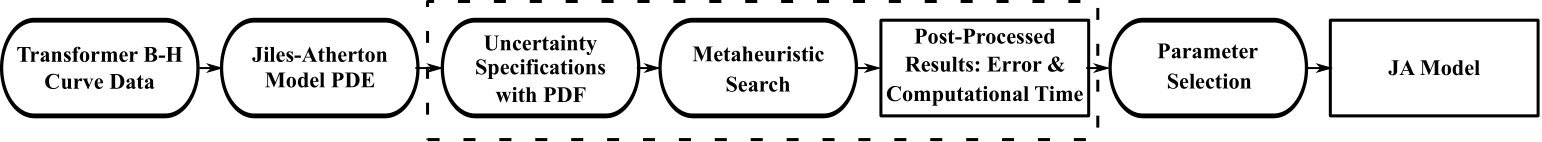}
	\caption{Overall block diagram of the proposed framework.}
	\label{fig:Flow_chart_general}
    \vspace{-0.3cm}
\end{figure*}

\subsection{Problem Statement}
\label{ssc:problem_statement}

Inrush current is a transient phenomenon that can harm the health and life expectancy of power transformers. To develop an efficient and reliable inrush minimization strategy it is necessary to model the precise dynamic response of the core material's magnetization. This nonlinear behaviour is represented by the B-H curve and is normally obtained by the JA hysteresis model. However, obtaining the JA model solution is a demanding task, determined by the resolution of the PDE in (\ref{eq:Inverse_JA}). This PDE is subjected to five parameters related to the physical properties of the core material. There parameters can be estimated through trial-and-error at the cost of a time-consuming process. To overcome these problems, metaheuristic-based search algorithms have been employed in the literature.

Additionally, proper initialization of JA parameters is essential to avoid convergence to a local minimum, non-convergence or high computational time. Generally, the JA parameter initialization is done by setting random limits and following a uniform distribution. Recent studies on basic simple functions have demonstrated that using different PDFs can enhance the accuracy and lower the computational time of metaheuristic-based algorithms \cite{Li2020}. However, this strategy is yet to be transferred to real industrial applications, such as transformer modelling for inrush current studies.

In this context, the aim of this research is to develop an uncertainty-aware JA parameter estimation framework, which enables (i) the assessment of data and expert knowledge with different uncertainty levels on JA parameter initialization and (ii) their propagation through metaheuristic-based algorithms for optimal parameter selection. This framework can guide the selection of the most suitable method to obtain the JA parameters, according to the data and knowledge of the engineer about the transformer. Therefore, this study can improve the modelling of the transformer core and the accuracy of transient studies, which can be the base for developing a proper and reliable inrush current minimization strategy.

\subsection{Proposed Model}
\label{ssc:Proposed_model}

Fig. \ref{fig:Flow_chart_general} shows the proposed method to identify the JA parameters based on the probabilistic initialization of parameters and metaheuristic optimization. To obtain the JA model, it is necessary to have the \texttt{B-H hysteresis loop data} of the transformer core, which is used to solve the \texttt{PDE determined by the JA model} [cf. (\ref{eq:Inverse_JA})]. Then, the JA parameters are initialized through different \texttt{uncertainty levels}, modelled with PDFs. The parameter initialization is followed by the execution of the chosen \texttt{metaheuristic-search} algorithm. The error and computational time obtained in each iteration are stored for \texttt{post-processing}. After the execution stage is finished, the optimal \texttt{parameter selection} for each metaheuristic algorithm and parameter initialization strategy are selected and used to \texttt{tune the JA model}.

\begin{algorithm*}[!t]
  \caption{Proposed JA parameter estimation algorithm.}
  \label{al:pseudocode1}
  \begin{algorithmic}
    \Require {\(B_{\text{data}}\), \(H_{\text{data}}\) } \Comment {B-H curve data}
    \State Select $dist(.) \gets$  \{$U(u, l)$\} $\vee$ \{\(N(\mu, \sigma)\)\} \Comment{Select initialization criteria, distribution $dist(.)$}
    \State Select $metaheuristics(params,f(.)) \gets$ \{GA $(params, f(.))$\} $\vee$ \{PSO $(params, f(.))\}$ $\vee$ \{DE $(params, f(.))$\} \newline 
 \Comment{Select metaheuristic algorithm, with corresponding input parameters and fitness function, as per (\ref{eq:erroe})}
    \For{\(i = 1:N\)} \Comment{Repeated sampling to infer mean statistics}
      \State $v_{\varepsilon}$ = 0 \Comment{Initialize error storage vector}
      \State \(v_{t} = 0\) \Comment{Initialize time storage vector}
     \State \(v_{\varepsilon_{it}} = 0\) \Comment{Initialize iteration error storage vector}
      \State \(M_{params} = 0\) \Comment{Initialize parameter matrix}
      \State \(it = 0\) \Comment{Initialize iteration counter }
      \State \(it_{\text{max}} = 10000\) \Comment{Initialize maximum number of iterations}
      \State \(\varepsilon_{it} = 100\) \Comment{Initialize error of the $it$ iteration}
      \State $\varepsilon_{limit}$ = 0 \Comment{Initialize acceptable error limit}
      \State \(t_{\text{count}} = 0\) \Comment{Initialize time counter}   
      \State \(\{M_{\text{s}}, a, \alpha, c, k\} \sim dist(.)\)  \Comment{Draw parameters from the selected distribution} 
      \While{\(\varepsilon_{it}>\varepsilon_{\text{limit}}\) \textbf{and} \(it < it_{\text{max}}\)} \Comment{Stop criteria for the metaheuristics}
        \State \(\{\text{$params$'}, \varepsilon_{it}, t_{\text{count}}\} \sim metaheuristics(params, f(.))\)  \Comment{Execute metaheuristics with fitness function $f(.)$ defined in (\ref{eq:erroe}) and $params$ defined for each algorithm in Subsection \ref{ssc:Metaheuristics}}
        \State \(it = it + 1\) \Comment{Update iteration}
        \State \(v_{\varepsilon_{it}}[i] = \varepsilon_{it}\) \Comment{Store the error of the iteration $it$}
        \If { \(it > 100\)} \Comment{Condition to update error limit}
        \State $\varepsilon_{limit}$= $v_{\varepsilon_{it}}[it-100]$ \Comment {Update error limit}
        \EndIf
      \EndWhile
      \State \(v_{\varepsilon}[i] = \varepsilon_{it}\) \Comment{Store optimized error}
      \State \(v_{t}[i] = t_{\text{count}}\) \Comment{Store time}  
      \State \(M_{\text{$params$}}[i, :] = \text{$params$'}\) \Comment{Store optimized $params$}
    \EndFor
    \State stats\_$v_{\varepsilon}$ = \texttt{infer\_stats}($v_\varepsilon$) \Comment{Infer, max. likelihood, 95$\%$ CI statistics, via \texttt{infer\_stats} function}    
    \State stats\_$v_{t}$ = \texttt{infer\_stats}($v_{t}$) \Comment{Infer, max. likelihood, 95$\%$ CI statistics via \texttt{infer\_stats} function}
    \Return {\(stats\_v_{\varepsilon}, stats\_v_{t}, M_{\text{params}}\)}\Comment{Distribution statistics for error and time, and optimized JA parameters}
  \end{algorithmic}
  \vspace{-2pt}
\end{algorithm*}

The three stages covered by a dashed box in Fig. \ref{fig:Flow_chart_general} are part of the computational process, which is further elaborated in the pseudocode presented in Algorithm \ref{al:pseudocode1}. After setting the B-H curve data as the algorithm input, the time starts counting and the parameters are initialized by the chosen strategy. Then, the chosen metaheuristic optimization method is executed. The selected minimization function is the root mean square error $\varepsilon$ between the measured and the calculated flux density in [\%]:

\begin{equation}\label{eq:erroe}
\varepsilon = \frac{1}{B_{\text{s}}}\sqrt{\sum_{i}^{N} \frac{\left(B_{\text{data}_i} - B_{\text{cal}_i}\right)^2}{N}} \text{x} 100
\end{equation} 

\noindent where $B_{\text{s}}$ is the saturation value of the flux density in [T], $i$ is the current iteration, $N$ is the number of flux density values, $B_{\text{data}_i}$ is the flux density value in [T] of the material at each iteration and $B_{\text{cal}_i}$ is the estimated flux density in [T] values of the material at each iteration.

As shown in the pseudocode in Algorithm \ref{al:pseudocode1}, the algorithm finishes when (i) it converges, \textit{i.e.} the error has not changed in the last 100 iterations or (ii) a maximum number of $it$ iterations have been reached ($it_\text{max}$ = 10000, in this study). Then, the error, computational time and JA optimized parameters are stored, and the same procedure is repeated $N = 10000$ times to obtain the error and computational time distributions.

The Algorithm \ref{al:pseudocode1} makes use of the \texttt{infer\_stats} function to analyze the error and computational time results. This function calculates the area under the curve of the distributions to ensure where the confidence interval (CI) of 95\% lays \cite{Aizpurua2019}.

\section{Case Studies}
\label{sec:case_study}

The presented framework is validated with two different materials and their B-H curves. The pseudocode presented in Algorithm \ref{al:pseudocode1} is executed with an Intel(R) Core(TM) i5-8265U CPU @ 1.60GHz processor. Material A is a H75-23 core used in medium voltage transformers and material B has been obtained from an open-source database of B-H curves \cite{Bhadra2022}. The JA parameter limits are set based on available information. For material A, the saturation is known and $M_{\text{s}}$ is not included in the estimation procedure. The rest of the material A parameters are set based on approximations and literature values \cite{Leite2004, Jiles1991_2}. The limits for material B are higher because previous analysis indicated that the parameter values for material B are higher than for material A \cite{Bhadra2022}. The chosen limits for both materials are presented in Table \ref{tab:limit}.

\begin{table}[h]
    \vspace{-0.2cm}
    \centering
    \caption{\textsc{Limits for JA parameters.}}
    \label{tab:limit}
    \begin{tabular}{|c|c|c|c|c|c|}
    \hline
    \multirow{2}{*}{\textbf{Parameter}} & \multicolumn{2}{c|}{\textbf{Limits Material A}}  & \multicolumn{2}{c|}{\textbf{Limits Material B}} & \multirow{2}{*}{\textbf{Unit}} \\ \cline{2-5}
                                    & \textbf{Lower}       & \textbf{Upper}      & \textbf{Lower}       & \textbf{Upper}      &                                \\ \hline
    \textit{\textbf{$M_{\text{s}}$}}                         & 1.52x10$^{6}$        & 1.52x10$^{6}$       & 9.35x10$^{5}$        & 11.42x10$^{5}$      & A/m                            \\ \hline
    \textbf{\textit{$a$}}                          & 1                    & 100                 & 1                    & 1000                & A/m                            \\  \hline
    \textbf{\textit{$\alpha$}}                      & 10$^{-6}$            & 10$^{-5}$           & 10$^{-6}$            & 10$^{-5}$           & {[}-{]}                        \\  \hline
    \textbf{\textit{$c$}}                          & 0.1                  & 0.9                 & 0.1                  & 0.9                 & {[}-{]}                        \\  \hline
    \textbf{\textit{$k$}}                          & 1                    & 100                 & 1                    & 5000                & A/m               \\ \hline
    
    \end{tabular}
    \vspace{-0.2cm} 
\end{table}

\subsection{Bruce-Force Base-Case}
\label{sc:Base_case}

In the base-case, the JA parameters are estimated by a brute-force algorithm. The algorithm sweeps the parameters between the selected limits and tries all the possible combinations. This case is used for comparison purposes.
\vspace{-0.1cm}
\subsection{Description of the Cases and Uncertainty}
\label{ssc:distribution_discussion}

The impact on JA parameter initialization with different PDFs has been analysed for the three algorithms and both materials. Firstly, the uniform parameter initialization is carried out, which models the case of least knowledge about the parameters, defined as follows:

\begin{equation} \label{eq:uniform}
f(x) = 
\begin{cases}
\frac{1}{u-l} & \text{for} \quad l \leq x \leq u \\
0 & \text{for} \quad x < l \quad \text{or} \quad x > u
\end{cases}
\end{equation}

\noindent where $l$ and $u$ are lower and upper boundaries respectively. These boundaries are given in Table \ref{tab:limit} for materials A and B.

Next, so as to model the level of uncertainty depending on available data and expert knowledge, the initialization of the JA parameters is done with the Gaussian PDFs, defined as:

\begin{equation}
\label{eq:GaussianPDF}
f(x) = \frac{1}{\sigma \sqrt{2\pi}} e^{\left(-\frac{(x-\mu)^2}{2\sigma^2}\right)}
\end{equation}

\noindent where $\mu$ is the mean value, and $\sigma$ is the standard deviation.

The impact of the parameter initialization with different PDF levels on the accuracy and computational time has been analysed through different uncertainty levels. The selected scenarios cover the parameter initialization with uniform distribution as $U(u,l)$ and Gaussian or normal distributions with $\sigma = \{1\%, 5\%, 10\% \}$ which can be formalized as $N(\mu, 0.01\mu)$, $N(\mu, 0.05\mu)$ and $N(\mu, 0.1\mu)$. Note that, the lower the percentage of the standard deviation, the bigger the expert knowledge. This process results in the probabilistic estimation of error and computational time for different search-based algorithms. Selected PDFs are further processed to obtain the maximum likelihood estimate and 95\% confidence interval values through the function $\texttt{infer\_stats}$ using numerical integration methods \cite{Aizpurua23}.

\section{Numerical Results}
\label{sec:numerical_results}

\subsection{Base-Case}
\label{sc:R_base_case}

\begin{table}[!t]
\vspace{-0.5cm}
\centering
\caption{\textsc{Base-Case results.}}
\label{tab:base}
\resizebox{0.45\textwidth}{!}{
\begin{tabular}{|c|c|c|c|}
\hline
\textbf{Parameter}            & \textbf{Material A} &  \textbf{Material B} &  \textbf{Unit}               \\ \hline
\textbf{$M_{\text{s}}$}       & 1519130          &     1067217         & A/m                         \\ \hline
\textbf{$a$}                  & 13               &           211                  & A/m                         \\ \hline
\textbf{$\alpha$}             & 1x10$^{-5}$      &       9x10$^{-5}$                   & [-]                           \\ \hline
\textbf{$c$} & 0.8    &       0.2               & [-]           \\ \hline
\textbf{$k$}                  & 24                      &     201                   & A/m                         \\ \hline
\textbf{Error}                & 2.33                     &      10.81                & \textbf{\%}                 \\ \hline
\textbf{Time}                 & 20.5                      &      167            & min                         \\ \hline
\end{tabular}
}
\end{table}

The best fitting JA parameters obtained with the brute-force algorithm for materials A and B are summarized in Table \ref{tab:base}, along with the error and the computational time. Note that the error obtained for material B is higher than for material A because of the search space. The sample size used to sweep the range of the parameters is bigger for material B, due to computational limitations, and therefore, the algorithm has a lower resolution. Moreover, the computational time is higher for material B owing to the parameter combinations that the brute-force algorithm needs to try.

\subsection{Uncertainty Assessment}
\label{sc:PDF_result}

\subsubsection{Material A}
\label{ssc:Material_A_results}

\begin{figure}[!t]
	\centering
	\includegraphics[width=0.8\columnwidth]{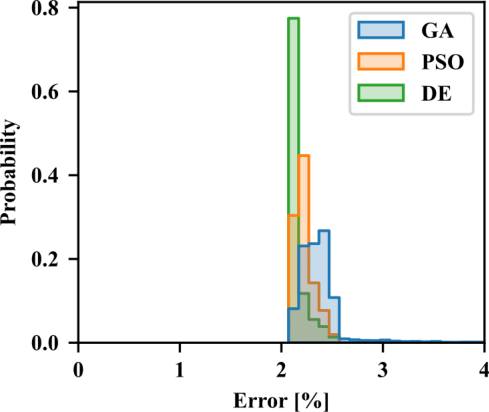}
	\caption{Error PDF for different algorithms with uniform parameter initialization for material A.}
	\label{fig:e_uniform_A}
\end{figure}

\begin{figure}[!t]
	\centering
	\includegraphics[width=0.8\columnwidth]{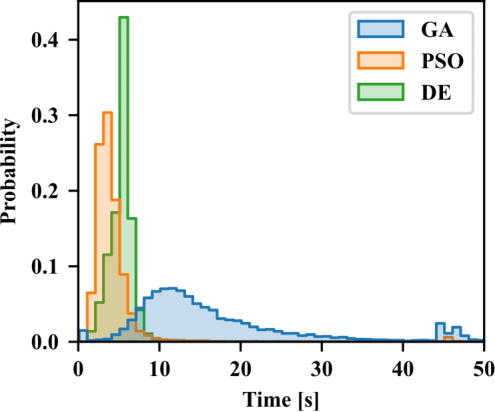}
	\caption{Computational time PDF for different algorithms with uniform parameter initialization for material A.}
	\label{fig:t_uniform_A}
    \vspace{-20pt}
\end{figure}

\begin{figure*}[!!ht]
\vspace{-5pt}
    \centering
    \begin{subfigure}[h]{0.32\textwidth}
        \centering
        \includegraphics[width=\textwidth]{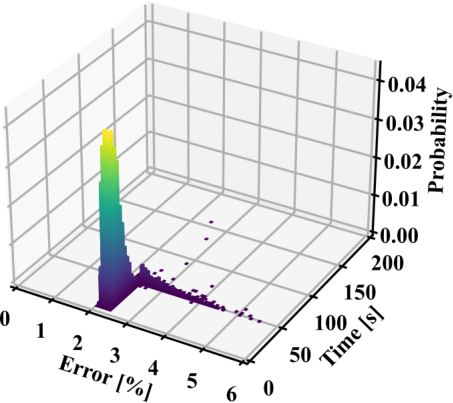}
        \caption{GA}
        \label{fig:GA_3D_A}
    \end{subfigure}
    \hfill
    \begin{subfigure}[h]{0.32\textwidth}
        \includegraphics[width=\textwidth]{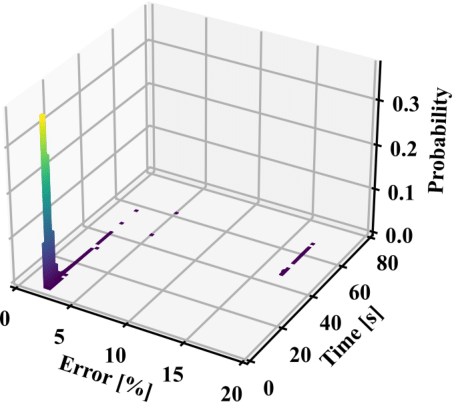}
        \caption{PSO}
        \label{fig:PSO_3D_A}
    \end{subfigure}
    \hfill
    \begin{subfigure}[h]{0.32\textwidth}
        \centering
        \includegraphics[width=\textwidth]{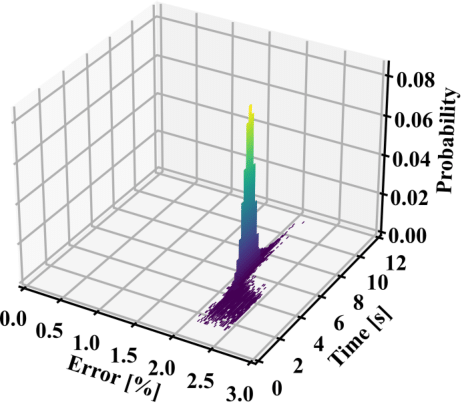}
        \caption{DE}
        \label{fig:DE_3D_A}
    \end{subfigure}
       \caption{Comparison of the probability distribution of error and computational time for material A.}
       \label{fig:3D_A}
\vspace{-10pt}
\end{figure*}

First, the JA parameter initialization has been carried out using a uniform distribution for the selected metaheuristic-based search algorithms. Fig. \ref{fig:e_uniform_A} shows that DE has the best accuracy results of 2.07$\%$. Moreover, further results displayed in Table \ref{tab:comparison_A} show that DE and PSO have a lower maximum likelihood error than GA. As for the computational time, Fig. \ref{fig:t_uniform_A} shows that PSO requires the least amount of computational time. Further results presented in Table \ref{tab:comparison_A} indicate that even though PSO is the fastest algorithm with a maximum likelihood computational time of 3s, it tends to converge to the local minimum instead of the global minimum, leading to a worst-case scenario of 17.26$\%$ error with a maximum computational time of 66s. This problem is evident in the 3D probability distribution plot between error and computational time shown in Fig. \ref{fig:PSO_3D_A}. On the other hand, GA is the slowest algorithm with a worst-case computational time of 154s, which can result in an error of 5.66$\%$. Moreover, the maximum likelihood results for GA are: computational time of 11s and an error of 2.49$\%$, which makes GA the less suitable algorithm for uniform parameter initialization. On the other hand, DE shows the best trade-off between accuracy and computational time, as demonstrated in the 3D plot shown in Fig. \ref{fig:DE_3D_A}, and is selected as the most suitable algorithm whenever no expert knowledge or previous data is available.

\begin{table*}[!b]
    \centering
 \caption{\textsc{Results for uniform and Gaussian PDF ($N(\mu, 0.05\mu)$) initialization with material A.}}
 \label{tab:comparison_A}
 \resizebox{\textwidth}{!}{
 \begin{tabular}{|c|c|c|c|c|c|c|c|} \cline{3-8}
 \multicolumn{1}{l}{} & & \multicolumn{2}{c|}{\textbf{Genetic Algorithm}} & \multicolumn{2}{c|}{\textbf{Particle Swarm Optimization}} & \multicolumn{2}{c|}{\textbf{Differential Evolution}} \\ \cline{3-8}
 \multicolumn{1}{l}{} & & \multicolumn{1}{c|}{\textbf{Uniform}} & \textbf{Normal 5\%} & \multicolumn{1}{c|}{\textbf{Uniform}} & \textbf{Normal 5\%} & \multicolumn{1}{c|}{\textbf{Uniform}} & \textbf{Normal 5\%} \\ \hline
 \multirow{3}{*}{\textbf{Error  [\%]}} & \textbf{Best/Worst Case} & 2.07/5.66 & 2.07/2.17 & 2.14/17.26 & 2.07/2.20 & 2.07/2.50 & 2.07/2.11 \\ 
 & \textbf{Max. Likelihood} & 2.49 & 2.12 & 2.15 & 2.08 & 2.15 & 2.15 \\
 & \textbf{95\% Upper/Lower} & 2.14/2.50 & 2.08/2.16 & 2.07/2.5 & 2.07/2.17 & 2.14/2.49 & 2.09/2.12 \\ \hline
 \multirow{3}{*}{\textbf{Time  [s]}} & \textbf{Best/Worst Case} & 1/154 & 1/2 & 1/66 & 1/7 & 1/11 & 1/6 \\
 & \textbf{Max. Likelihood} & 11 & 1 & 3 & 1 & 6 & 2 \\
 & \textbf{95\% Upper/Lower} & 5/53 & 0/2 & 1/15 & 0/6 & 1/9 & 0/6 \\ \hline
 \multirow{5}{*}{\textbf{Parameter}} & \textbf{\textit{$\bm{M_\text{s}}$}  [A/m]} & 1519130 & 1519130 & 1519130 & 1519130 & 1519130 & 1519130 \\
 & \textbf{\textit{a} [A/m]} & 12.95 & 12.89 & 12.83 & 13.52 & 12.82 & 13.38 \\
 & $\bm{\alpha}$ & 1.00x$10^{-5}$ & 1.01x10$^{-5}$ & 1.04x10$^{-5}$ & 1.16x10$^{-5}$ & 0.83x10$^{-5}$ & 1.12x10$^{-5}$ \\
 & \textbf{\textit{k}  [A/m]} & 24.06 & 23.75 & 23.95 & 25.10 & 23.49 & 25.32 \\
 & \textbf{\textit{c}} & 0.9 & 0.9 & 0.9 & 0.9 & 0.9 & 0.9 \\ \hline
 \end{tabular}}
 \end{table*}

\begin{figure}[!!t]
	\centering
	\includegraphics[width=0.925\columnwidth]{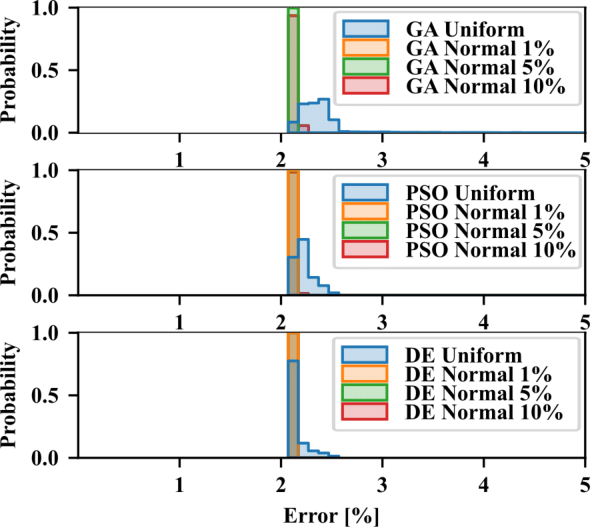}
	\caption{Comparison of the error between the different parameter initialization settings for material A.}
	\label{fig:e_comparison_dist_A}
    \vspace{-20pt} 
\end{figure} 

\begin{figure}[!t]
	\centering
	\includegraphics[width=0.925\columnwidth]{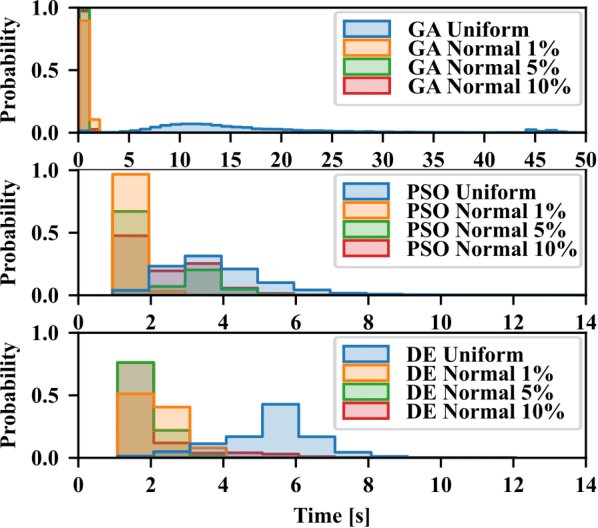}
	\caption{Comparison of the computational time between the different parameter initialization settings for material A.}
	\label{fig:t_comparison_dist_A}
    \vspace{-20pt}
\end{figure}    

 \begin{figure}[ht!]
	\centering
	\includegraphics[width=0.8\columnwidth]{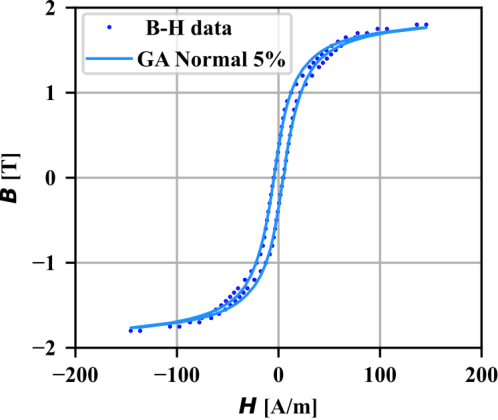}
	\caption{Comparison between the measured and the estimated B-H curves for material A.}
	\label{fig:BH_results_A}
    \vspace{-20pt} 
\end{figure} 

As illustrated in Fig. \ref{fig:e_comparison_dist_A}, when a Gaussian PDF is used for the parameter initialization, the accuracy is highly improved for all algorithms. Therefore, even the local minimum problem detected with PSO is solved. As for the computational time, the distribution comparison in Fig. \ref{fig:t_comparison_dist_A} shows the reduction in the execution time as the uncertainty decreases. Note that the scale of the subplots is different for improved visualization. The best trade-off between accuracy and computational time for each algorithm is found when the JA parameter initialization is done with a 5$\%$ Gaussian PDF. The results obtained for these cases are detailed in Table \ref{tab:comparison_A} along with the results obtained with the uniform parameter initialization strategy. In all scenarios, the maximum likelihood convergence speed for Gaussian PDF parameter initialization is between 1s and 2s. The best compromise between the accuracy and the computational time is given by GA. Hence, for material A, GA is the most suitable algorithm if expert knowledge or previous information is available. To verify the validity of the results, the B-H curve is obtained from the PDE in (\ref{eq:Inverse_JA}) using the most accurate case, which is GA Gaussian PDF with 5\% uncertainty. Fig. \ref{fig:BH_results_A} shows the agreement between the estimated B-H curves and the real data for material A.

\subsubsection{Material B}
\label{ssc:Material_B_results}

\begin{figure}[!t]
	\centering
	\includegraphics[width=0.8\columnwidth]{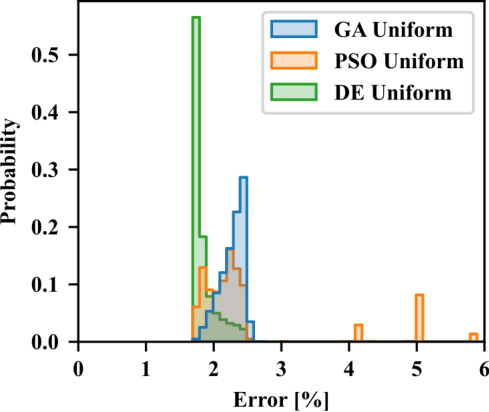}
	\caption{Error PDF for different algorithms with uniform parameter initialization for material B.}
	\label{fig:e_comparison_dist_B}
    \vspace{-20pt}
\end{figure}

\begin{figure}[!ht]
	\centering
	\includegraphics[width=0.85\columnwidth]{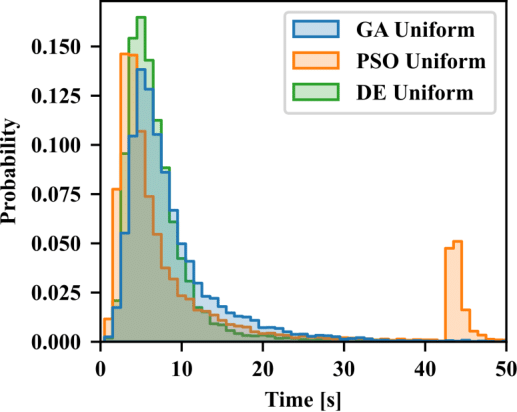}
	\caption{Computational time PDF for different algorithms with uniform parameter initialization for material B.}
	\label{fig:t_comparison_dist_B}
 \vspace{-20pt}
\end{figure}

\begin{figure*}[!!ht]
    \vspace{-5pt}
    \centering
    \begin{subfigure}[b]{0.32\textwidth}
        \centering
        \includegraphics[width=\textwidth]{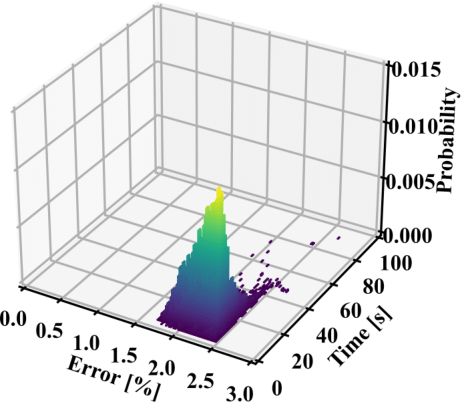}
        \caption{GA}
        \label{fig:GA_3D_B}
    \end{subfigure}
    \hfill
    \begin{subfigure}[b]{0.32\textwidth}
        \includegraphics[width=\textwidth]{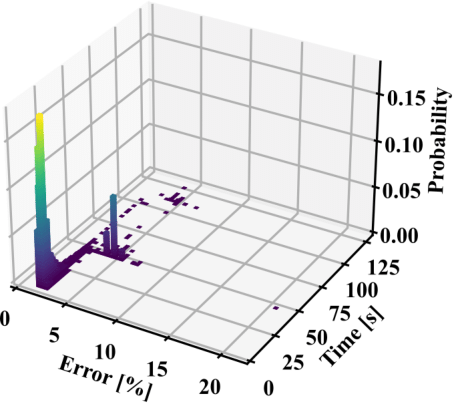}
        \caption{PSO}
        \label{fig:PSO_3D_B}
    \end{subfigure}
    \hfill
    \begin{subfigure}[b]{0.32\textwidth}
        \centering
        \includegraphics[width=\textwidth]{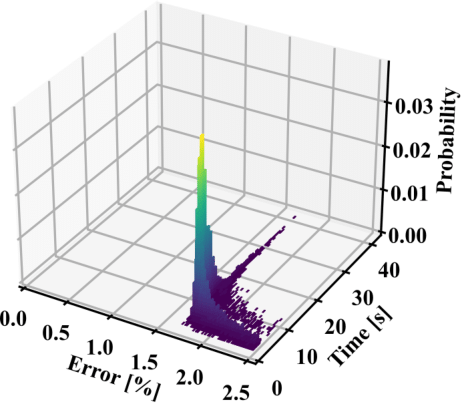}
        \caption{DE}
        \label{fig:DE_3D_B}
    \end{subfigure}
       \caption{Comparison of the probability distribution of error and computational time for material B.}
       \label{fig:3D_B}
    \vspace{-10pt}
\end{figure*}

In the case of the uniform PDF parameter initialization, Fig. \ref{fig:e_comparison_dist_B} shows that DE has the highest accuracy, with a maximum likelihood error of 1.69$\%$. PSO shows the worst accuracy with a maximum likelihood error of 5.01\%, as indicated in Table \ref{tab:comparison_B}. Moreover, PSO results show that this algorithm can get stuck in a local minimum with a worst-case error of 21.04\%. As for the computational time, Fig. \ref{fig:t_comparison_dist_B} shows that even though PSO has the highest probability of converging faster, it can be the slowest algorithm in case of converging to a local minimum. The worst-case computational time of PSO is 124s. The problem of local minima is better illustrated in the 3D plot in Fig. \ref{fig:PSO_3D_B}, where the PDFs of error and computational time are shown. On the other hand, even though the worst-case error obtained with GA is 2.72\%, the maximum computational time is 90s. Therefore, for material B, DE is also the algorithm with the best trade-off between error and computational time for a uniform PDF parameter initialization, as confirmed in Fig. \ref{fig:DE_3D_B}. 

\begin{figure}[!ht]
	\centering
	\includegraphics[width=0.925\columnwidth]{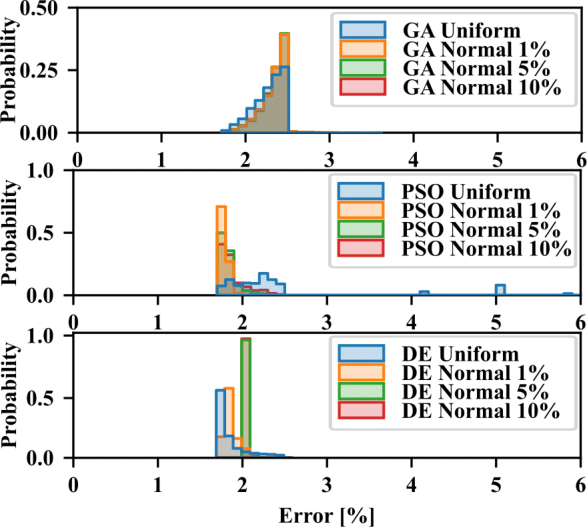}
	\caption{Comparison of the error between the different parameter initialization settings for material B.}
	\label{fig:e_dist_B}
    \vspace{-20pt}
\end{figure}

\begin{figure}[!ht]
	\centering
	\includegraphics[width=0.925\columnwidth]{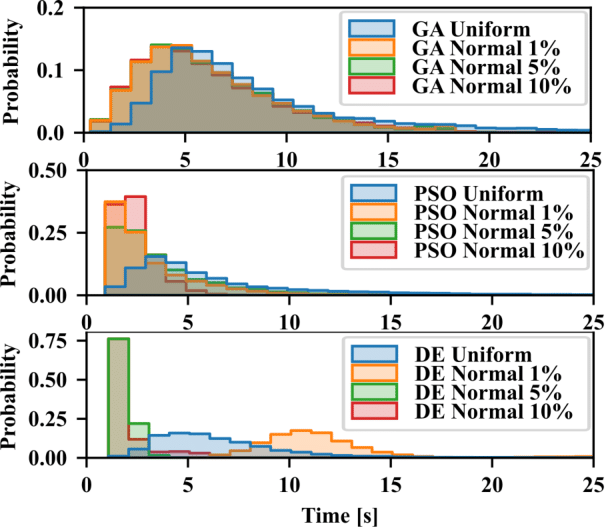}
	\caption{Comparison of the computational time between the different parameter initialization settings for material B.}
	\label{fig:t_dist_B}
    \vspace{-20pt} 

\end{figure}

Fig. \ref{fig:e_dist_B} shows the comparison between the error obtained for the different parameter initialization strategies. The results obtained for PSO are considerably improved when using the Gaussian PDF. Nonetheless, the error obtained by GA is not affected by the parameter initialization strategy, and DE results are worsened. However, as shown in Fig. \ref{fig:t_dist_B}, the computational time for DE improves for the 5\% and 10\% uncertainty cases and GA is not affected by the initialization strategy. On the other hand, PSO computational speed is reduced for the Gaussian PDF parameter initialization strategy. For a more detailed comparison, the results obtained with the 5$\%$ Gaussian PDF parameter initialization for all the algorithms are summarized in Table \ref{tab:comparison_B}. By analyzing all the results, for material B, PSO is found to be the algorithm with the best performance when a Gaussian PDF-based parameter initialization strategy is adopted.

\begin{table*}[!ht]
\centering
\caption{\textsc{Results for uniform and Gaussian PDF ($N(\mu, 0.05\mu)$) initialization with material B.}}
\label{tab:comparison_B}
\vspace{5pt}
\resizebox{\textwidth}{!}{
\begin{tabular}{|c|c|c|c|c|c|c|c|}
\cline{3-8}
 \multicolumn{1}{l}{} & & \multicolumn{2}{c|}{\textbf{Genetic Algorithm}} & \multicolumn{2}{c|}{\textbf{Particle Swarm Optimization}} & \multicolumn{2}{c|}{\textbf{Differential Evolution}} \\
\cline{3-8}
\multicolumn{1}{l}{} & & \multicolumn{1}{c|}{\textbf{Uniform}} & \textbf{Normal 5\%} & \multicolumn{1}{c|}{\textbf{Uniform}} & \textbf{Normal 5\%} & \multicolumn{1}{c|}{\textbf{Uniform}} & \textbf{Normal 5\%} \\ \hline
\multirow{3}{*}{\textbf{Error [\%]}} 
& \textbf{Best/Worst Case} & 1.71/2.72 & 1.75/3.56 & 1.70/21.04 & 1.74/2.19 & 1.69/2.50 & 2.07/2.11 \\ 
& \textbf{Most Likelihood} & 2.47 & 2.49 & 5.01 & 1.75 & 1.7 & 2.08 \\ 
& \textbf{95\% Upper/Lower} & 1.70/2.50 & 1.93/2.51 & 1.77/8.27 & 1.74/2.48 & 1.69/2.49 & 2.06/2.11 \\
\hline

\multirow{3}{*}{\textbf{Time [s]}} 
& \textbf{Best/Worst Case} & 1/90 & 1/29 & 1/124 & 1/31 & 1/41 & 1/6 \\  
& \textbf{Max. Likelihood} & 5 & 4 & 3 & 2 & 5 & 2 \\ 
& \textbf{95\% Upper/Lower} & 2/42 & 1/19 & 1/47 & 0/20 & 2/18 & 0/6 \\
\hline

\multirow{5}{*}{\textbf{Parameter}} 
& \textbf{\textit{$\bm{M_\text{s}}$} [A/m]} & 1015196 & 1018404 & 1013978 & 1021701 & 1015781 & 1015781 \\ 
& \textbf{\textit{a}} & 63.94 & 62.82 & 62.53 & 69.01 & 63.89 & 63.99 \\ 
& $\bm{\alpha}$ & 8.46x10$^{-5}$ & 9.78x10$^{-5}$ & 8.00x10$^{-5}$ & 10x10$^{-5}$ & 8.07x10$^{-5}$ & 8.22x10$^{-5}$ \\ 
& \textbf{\textit{k} [A/m]} & 274.01 & 378.16 & 272.49 & 302.51 & 269.72 & 264.86 \\ 
& \textbf{\textit{c}} & 0.48 & 0.64 & 0.46 & 0.54 & 0.46 & 0.45 \\ 
\hline
\end{tabular}
}
\vspace{-10pt}
\end{table*}

\begin{figure}[ht!]
	\centering
	\includegraphics[width=0.8\columnwidth]{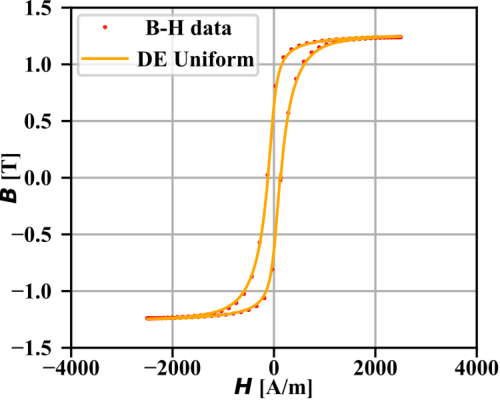}
	\caption{Comparison between the measured and the estimated B-H curves for material B.}
	\label{fig:BH_results_B}
    \vspace{-20pt} 
\end{figure}

The error and computational time of all the cases are summarized in Table \ref{tab:comparison_B}. To verify the validity of the results, the B-H curve is obtained from the PDE presented in (\ref{eq:Inverse_JA}) using the most accurate case, which for material B is DE uniform PDF. Fig. \ref{fig:BH_results_B} shows the agreement between the estimated B-H curves and the real data for material B. 

\subsection{Analysis of Results}

\subsubsection{Accuracy \& Computational Complexity}

First, to verify the improvement of the JA parameter estimation with metaheuristics, a base-case based on brute-force is analysed for both materials. Metaheuristic-based search algorithms can reduce the error from 2.33\%, obtained with brute-force, to 2.07\% for material A. The error reduction in material B is more significant, from 10.81\%, obtained with brute-force, to 1.69\%. As shown in Table~\ref{tab:base}, the search area for material B is greater than material A. Due to memory constraints, the resolution for material B is decreased and hence, the brute-force algorithm is not able to further reduce the error. The computational time is reduced from 20.5 minutes, in the brute-force case, to below 1 minute when the parameter initialization is carried out by a Gaussian PDF. In the case of material B, the brute-force algorithm lasts 167 minutes because the search space is higher. The computational time is shown to be reduced to below 1 minute when a Gaussian PDF is used.

\subsubsection{Parameter Initialization Strategies}

The most significant issue regarding uniform parameter initialization is the problem of convergence to a local minimum encountered with PSO. Even though this is outside the 95\% confidence interval, as shown in Tables~\ref{tab:comparison_A} and \ref{tab:comparison_B}, the results for material B show that the upper 95\% quantile for the error and computational time are 8.27\% and 47s, respectively. In the case of GA, the 95\% confidence interval shows good performance for material A. However, for material B, the computational time 95\% upper confidence interval has a value of 42s. Therefore, DE is found to have the best performance in terms of accuracy and computational time for both materials when using uniform distribution. This is visualized in the 3D plots shown in Figs. \ref{fig:3D_A} and \ref{fig:3D_B}, where the error and computational time distributions for uniform parameter initialization are shown.

The parameter initialization with Gaussian PDFs has solved the local minimum problem for PSO, and results show that the accuracy and computational time of this algorithm are improved for both materials. However, for GA and DE, the utilization of PDFs for parameter initialization has yielded different results for each material. In the case of GA, the performance of the algorithm is shown to improve for material A, as shown in Figs. \ref{fig:e_comparison_dist_A} and \ref{fig:t_comparison_dist_A}. However, for material B, GA has obtained similar results for all the parameter initialization strategies. This is in accordance with the study presented in Section \ref{sec:Intro} about the impact of different initialization strategies, which concludes that GA has lower sensitivity than PSO to the parameter initialization strategy. Conversely, the performance of DE is improved with Gaussian PDF parameter initialization in the case of material A and obtains the best performance among the tested search algorithms. However, for material B, the error distribution shows a higher probability of decreasing the error for random uniform parameter initialization than for Gaussian PDFs. Therefore, these results demonstrate that the algorithm with the best performance for Gaussian PDF parameter initialization for both materials is PSO.

\section{Discussion}
\label{sec:discussion}

This research presents a framework to assess the initialization of JA parameters for their estimation using metaheuristic search algorithms. The parameter initialization strategy is determined by different uncertainty levels depending on previous data and expert knowledge. Results show that the use of PDFs for parameter initialization improves the accuracy and computational time of the metaheuristic search algorithm with respect to classical random uniform initialization strategies. However, before drawing definitive conclusions about the suitability of the analysed metaheuristic algorithms for modelling the transformer core, further work is necessary to test the approach for analysing electromagnetic transients, such as inrush current, in transformers.

The transformer modelling process is a complex task, which requires the coordination of different models and information sources, e.g. duality-based transformer model \cite{Chiesa2010_2}. In this context, the JA parameter estimation is an offline process carried out once during the design stage. Even in time-sensitive applications like transformer’s CB auto-reclosing, related to inrush current, the computational time needed for JA parameter estimation is not significant \cite{Sanati2022}. However, the transformer design process consists of various demanding stages, such as optimizing the size and losses, while keeping the temperature rise below safe values. The accuracy in the calculation of losses (and temperature rise) is also crucial. In parallel, other design requirements must also be met, including clamp forces or transformer insulation codes.

In this context, cost optimization is essential, and in a competitive market, fast and reliable design processes are key. Therefore, any reduction in the execution time of these stages holds potential economic benefits for the company and alleviates the workload for engineers.

\section{Conclusions}
\label{sec:Conclusions}

This study presents a framework to assess the JA parameter initialization including different levels of uncertainty present in available data and expert knowledge through metaheuristic-based search algorithms. The proposed approach uses uncertainty awareness through PDFs and their propagation in three different metaheuristic algorithms including GA, PSO and DE. The obtained accuracy and computational time results are compared and analysed for two different transformer core materials. Results show that the accuracy and computational time of the JA parameter estimation for both materials have been improved when a Normal PDF initialization strategy is used instead of the uniform parameter initialization.

The proposed approach eases the decision-making process for JA-based hysteresis modelling approach for novel and experienced engineers alike. A novel engineer with limited field experience may benefit from the proposed framework through the selection of the most accurate and computationally efficient metaheuristic-search algorithm using the random uniform parameter initialization. Obtained results show that DE metaheuristic algorithm is most appropriate among the analysed algorithms for JA parameter estimation scenarios with lack of available information. For experienced engineers, the proposed parameter initialization approach is shown to improve accuracy and computational time with the selection of a Normal PDF. Therefore, the framework presented in this research may benefit end users with different transformer and field experiences.

\bibliographystyle{IEEEtran}

\bibliography{IEEEabrv,main_document}

\newpage

\begin{IEEEbiography}[{\includegraphics[width=1in,height=1.25in,clip,keepaspectratio]{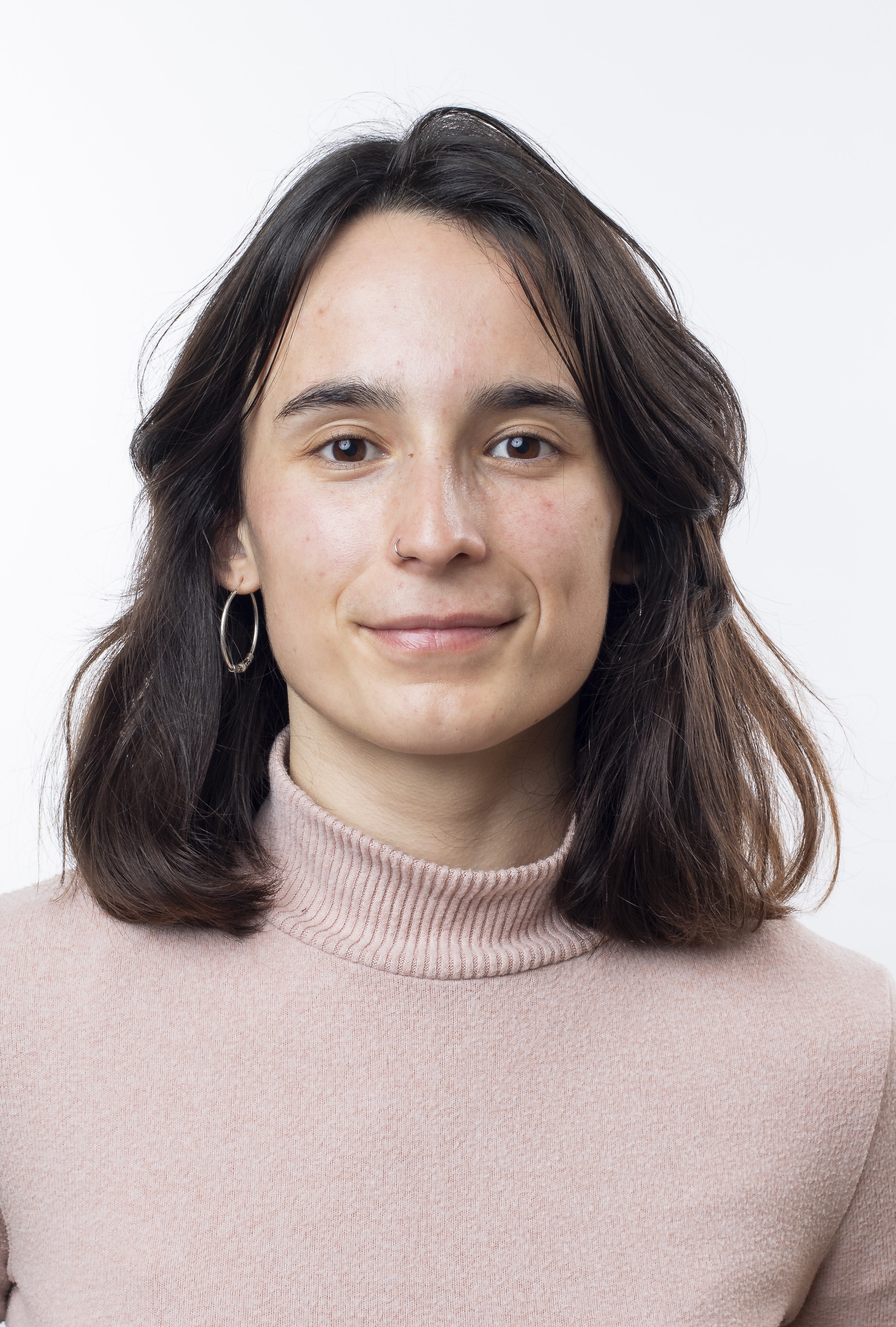}}]
{Jone Ugarte Valdivielso} received her B.Sc degree in Energy Engineering from Mondragon University (MU) in 2019 and her M.Sc. in Electrical Power Systems and High Voltage from Aalborg University (AAU) in 2021. Since 2020, she has been with the Electronics and Computer Science Department at MU. Since 2023, she is pursuing her PhD in inrush current minimization on power transformers. Her research interests include medium- and high-voltage technology in electrical power systems.
\end{IEEEbiography}

\vskip -2\baselineskip plus -1fil

\begin{IEEEbiography}[{\includegraphics[width=1in,height=1.25in,clip,keepaspectratio]{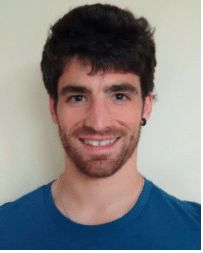}}]
{Jose I. Aizpurua (M'17, SM'22)} is a Lecturer \& Researcher at Mondragon University (MU) and Ikerbasque Research Fellow. He received his Eng., M.Sc., and Ph.D. degrees from MU in 2010, 2012, and 2015, respectively. He was a Visiting Researcher at the Univ. of Hull (2014), and Research Associate at the Univ. of Strathclyde in Glasgow (02/2015-12/2018). Dr Aizpurua's research interests include the development of intelligent solutions for the power and energy sector.
\end{IEEEbiography}

\vskip -2\baselineskip plus -1fil

\begin{IEEEbiography}[{\includegraphics[width=1in,height=1.25in,clip,keepaspectratio]{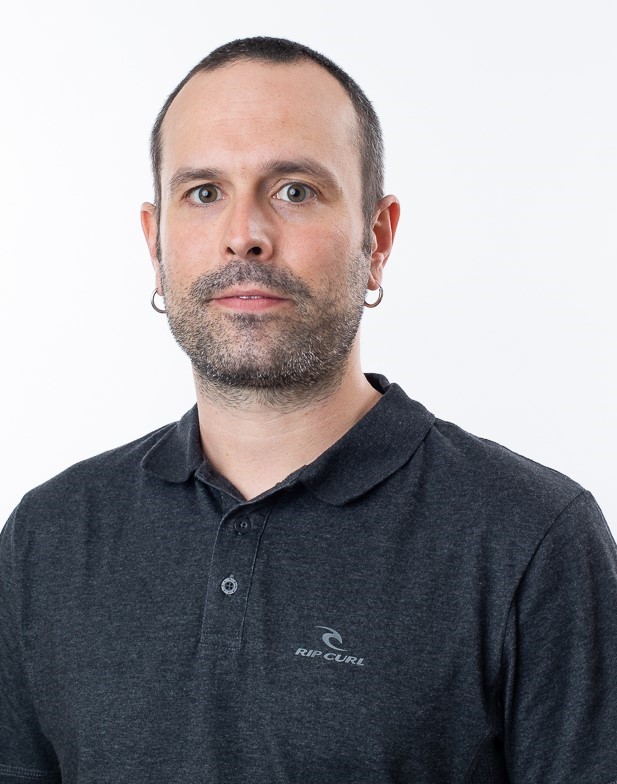}}]
{Manex Barrenetxea} received his M.Sc. and Ph.D. degrees in electrical engineering from Mondragon University, Arrasate, Spain, in 2011 and 2016, respectively. Since 2016, he works as a lecturer and researcher in the Electronics and Computer Science Department of Mondragon University. Since 2020, he is the Head of the Medium Voltage Laboratory of Mondragon University. He has coauthored the open-access book “Power electronic converter design handbook”. His research interests include power electronics and medium voltage applications.
\end{IEEEbiography}

\end{document}